\documentclass[reprint, longbibliography, noeprint, superscriptaddress]{revtex4-1}

\usepackage[utf8]{inputenc}

\usepackage[lining,semibold]{libertine} 
\usepackage[libertine, cmintegrals, bigdelims, vvarbb]{newtxmath}

\usepackage{amsmath}
\usepackage{amsfonts}
\usepackage{mathrsfs}
\usepackage{gensymb}
\usepackage{bbm}
\usepackage{dsfont}

\usepackage{kbordermatrix}
\renewcommand{\kbldelim}{(}
\renewcommand{\kbrdelim}{)}

\usepackage{chemformula}
\usepackage[caption=false]{subfig}
\usepackage{chemfig}
\usepackage[version=3]{mhchem}

\usepackage{tikz}
\usetikzlibrary{matrix,positioning,decorations.pathreplacing}

\usepackage{soul}
\usepackage{xcolor}

\usepackage{scalerel} 
\usepackage{comment}

\usepackage{blkarray}

\definecolor{webgreen}{rgb}{0,.5,0}
\definecolor{webbrown}{rgb}{.6,0,0}
\definecolor{grigio}{rgb}{.85,.85,.85} 
\definecolor{RoyalBlue}{rgb}{0.0, 0.14, 0.4}
\definecolor{skyblue1}{rgb}{0.45,0.62,0.81}
\definecolor{skyblue2}{rgb}{0.2,0.39,0.64}
\definecolor{skyblue3}{rgb}{0.13,0.29,0.53}
\definecolor{scarlet1}{rgb}{0.93,0.16,0.16}
\definecolor{scarlet2}{rgb}{0.8,0,0}
\definecolor{scarlet3}{rgb}{0.64,0,0}

\definecolor{g}{gray}{0.50}

\usepackage{hyperref}
\hypersetup{%
    colorlinks=true, linktocpage=true, pdfstartpage=1, pdfstartview=FitV,%
    breaklinks=true, pdfpagemode=UseNone, pageanchor=true, pdfpagemode=UseOutlines,%
    plainpages=false, bookmarksnumbered, bookmarksopen=true, bookmarksopenlevel=1,%
    hypertexnames=true, pdfhighlight=/O,
    urlcolor=webbrown, linkcolor=RoyalBlue, citecolor=webgreen, 
    pdftitle={},%
    pdfauthor={Francesco Avanzini},%
    pdfsubject={},%
    pdfkeywords={},%
    pdfcreator={pdfLaTeX},%
    pdfproducer={LaTeX REVTeX}%
}

\usepackage{cleveref}

\newcommand{\rem}[1]{{\color{black}#1}}
\newcommand{\cmark}[1]{{\color{black}#1}}

\newcommand{\remark}{\textit{\rem{Remark.} }}

\newcommand{\assumptionrf}{\textbf{{Reaction Fluxes.}} }

\newcommand{\implementationM}{\textbf{{Implementation of the method.}} }

\newcommand{\effp}[1]{\hat{#1}}

\newcommand{\abs}[1]{|#1|}

\newcommand{\dt}{\mathrm{d}_t}
\newcommand{\pt}{\partial_t}
\newcommand{\dd}{\mathrm d}

\newcommand{\dk}{{\mathbb 1}}

\newcommand{\pos}{\boldsymbol{r}}

\newcommand{\ds}{\boldsymbol{\nabla}}

\newcommand{\setspe}{\mathcal S}

\newcommand{\setspex}{X}
\newcommand{\setspexi}{X_I}
\newcommand{\setspexd}{X_D}

\newcommand{\setspey}{Y}

\newcommand{\setcom}{\mathcal K}

\newcommand{\setrct}{\mathcal R}

\newcommand{\spe}{i}
\newcommand{\spee}{j}

\newcommand{\spex}{x}

\newcommand{\spey}{y}

\newcommand{\rct}{\rho}

\newcommand{\com}{\kappa}
\newcommand{\comrct}[1]{\kappa(#1\rct)}

\newcommand{\chem}{Z_{\spe}}
\newcommand{\chemx}{Z_{\spex}}
\newcommand{\chemy}{Z_{\spey}}

\newcommand{\comp}[1]{V_{\com#1}}

\newcommand{\stoc}[2]{\nu_{#2, #1\rct}}

\newcommand{\stocx}[1]{{\nu}_{\spex, #1\rct}}
\newcommand{\stocy}[1]{{\nu}_{\spey, #1\rct}}

\newcommand{\matS}{\mathbb{S}}

\newcommand{\matSX}{\mathbb{S}_{\setspex}}

\newcommand{\matSi}[1]{{S}_{\spe,#1\rct}}
\newcommand{\matSx}[1]{{S}_{\spex,#1\rct}}
\newcommand{\matSy}[1]{{S}_{\spey,#1\rct}}
\newcommand{\matSyy}[1]{{S}_{\spey,#1\rct'}}

\newcommand{\tsq}{\alpha_{|\rct|}}
\newcommand{\tsqq}{\alpha_{|\pth|}}

\newcommand{\matI}{\partial}
\newcommand{\matIe}{\partial_{\com,\rct}}
\newcommand{\matC}{\mathbb{\Gamma}}
\newcommand{\matCe}[1]{\Gamma_{\spex,\com#1}}

\newcommand{\deff}{\delta}

\newcommand{\setpth}{\mathcal{E}}

\newcommand{\pth}{\varepsilon}
\newcommand{\pthrct}[1]{\varepsilon(#1\rct)}
\newcommand{\setrctpth}[1]{\setrct_{#1\pth}}

\newcommand{\stocxe}[1]{{\nu}_{\spex, #1\pth}}

\newcommand{\matSxe}[1]{{S}_{\spex,#1\pth}}
\newcommand{\matSxeI}[1]{\overline{S}_{#1\pth, \spex}}

\newcommand{\cyclev}{\boldsymbol{\phi}}
\newcommand{\cycle}{\phi_{\rct}}
\newcommand{\cycleev}{\hat{\boldsymbol{\phi}}}
\newcommand{\cyclee}{\hat{\phi}_{\pth}}

\newcommand{\conc}{c_\spe(\pos)}

\newcommand{\conv}{\boldsymbol{c}(\pos)}
\newcommand{\convv}{\boldsymbol{c}}

\newcommand{\convveq}{\boldsymbol{c}_{\mathrm{eq}}}
\newcommand{\convss}{\boldsymbol{c}_{\mathrm{ss}}(\pos)}
\newcommand{\convvss}{\boldsymbol{c}_{\mathrm{ss}}}

\newcommand{\convtot}{\boldsymbol{c}_{\mathrm{tot}}(\pos)}

\newcommand{\convssh}{\boldsymbol{c}_{\mathrm{ss}}^{\mathrm{h}}}

\newcommand{\conx}{c_\spex^{}(\pos)}

\newcommand{\conyy}{c_\spey^{}}

\newcommand{\dcurr}{\boldsymbol{J}_{\spe}(\conv)}

\newcommand{\dcurrss}{\boldsymbol{J}_{\spe}(\convss)}

\newcommand{\matO}{\mathbb{O}(\conv)}
\newcommand{\matOij}{O_{\spe,\spee}(\conv)}

\newcommand{\rcurr}[1]{j_{#1\rct}(\conv)}

\newcommand{\rcurrss}[1]{j_{#1\rct}(\convss)}

\newcommand{\rcurrssh}[1]{j_{#1\rct}(\convssh)}

\newcommand{\rcurravg}[1]{\overline{j}_{#1\rct}}

\newcommand{\rflux}[1]{\omega_{#1\rct}(\conv)}
\newcommand{\rfluxx}{\omega^{\mathrm{in}}(\conv, \{\stoc{}{\spex}\})}

\newcommand{\kconst}[1]{k_{#1\rct}}

\newcommand{\rfluxxx}{\omega^{\mathrm{in}}(\conv, \{\stoc{}{\spex}\},  \{\stoc{-}{\spex}\})}
\newcommand{\rfluxxy}[1]{\omega_{#1\rct}^{\mathrm{ch}}(\{\conyy\})}

\newcommand{\rfluxxpth}[1]{\omega^{\mathrm{in}}_{\pthrct{#1}}(\conv)}
\newcommand{\rfluxxcom}[1]{\omega^{\mathrm{in}}_{\comrct{#1}}(\conv)}
\newcommand{\rfluxxcomm}{\omega^{\mathrm{in}}_{\com}(\conv)}

\newcommand{\rfluxar}[1]{\omega_{#1\rct}^{\mathrm{ar}} (\conv)}
\newcommand{\rfluxts}[1]{\omega_{#1\rct}^{\mathrm{ts}} (\conv)}

\newcommand{\rfluxex}[1]{\omega_{#1} (\conv)}

\newcommand{\srr}{s_{|\rct|}(\conv)}

\newcommand{\ecurr}{I_{\spe}(\pos)}

\newcommand{\ecurry}{I_{\spey}(\pos)}

\newcommand{\pcurr}[1]{\effp{j}_{#1\pth}(\conv)}

\newcommand{\pflux}[1]{\effp{\omega}_{#1\pth}(\conv)}

\newcommand{\AAAp}{A_{|\rct|}}

\newcommand{\aap}[1]{a_{#1\pth}(\conv)}
\newcommand{\bbp}[1]{b_{#1\pth}}

\newcommand{\fd}[1]{{\delta#1}/{\delta\conc}}

\newcommand{\free}{F[\convv]}
\newcommand{\freess}{F[\convvss]}

\newcommand{\cp}{\mu_\spe(\conv)}

\newcommand{\cpg}[1]{\mu_{#1}(\conv)}

\newcommand{\cpgssh}[1]{\mu_{#1}(\convssh)}

\newcommand{\cpss}{\mu_\spe(\convss)}

\newcommand{\cpx}{\mu_\spex(\conv)}

\newcommand{\stcpy}{\mu_\spey^\circ}

\newcommand{\cpy}{\mu_\spey(\conv)}
\newcommand{\cpyy}{\mu_\spey}

\newcommand{\eprdff}{\dot{\Sigma}_{\mathrm{dff}}[\convv]}

\newcommand{\eprrct}{\dot{\Sigma}_{\mathrm{rct}}[\convv]}

\newcommand{\eprrcte}{\effp{\dot{\Pi}}[\convv]}

\newcommand{\eprdffeq}{\dot{\Sigma}_{\mathrm{dff}}[\convveq]}
\newcommand{\eprrcteq}{\dot{\Sigma}_{\mathrm{rct}}[\convveq]}

\newcommand{\eprdffss}{\dot{\Sigma}_{\mathrm{dff}}[\convvss]}
\newcommand{\eprrctss}{\dot{\Sigma}_{\mathrm{rct}}[\convvss]}

\newcommand{\chwrk}{\dot{W}_{\mathrm{chm}}[\convv]}
\newcommand{\chwrkeq}{\dot{W}_{\mathrm{chm}}[\convveq]}
\newcommand{\chwrkss}{\dot{W}_{\mathrm{chm}}[\convvss]}

\newcommand{\kinfree}{\effp{F}[\convv]}
\newcommand{\kineprrct}{\effp{\dot{\Pi}}[\convv]}
\newcommand{\kineprrctss}{\effp{\dot{\Pi}}[\convvss]}

\newcommand{\shx}{\Delta_{\spex}}

\newcommand{\freecb}{\hat{F}[\convv]}

\newcommand{\cpxssh}{\mu_\spex(\convssh)}

\newcommand{\eprrctc}{\dot{\Pi}[\convv]}

\newcommand{\rna}{a}
\newcommand{\rnb}{b}

\newcommand{\pathway}{{lumped reaction}}
\newcommand{\pathways}{{lumped reactions}}

\newcommand{\Pathways}{{Lumped reactions}}

\newcommand{\method}{method}

\makeatletter
\def\maketag@@@#1{\hbox{\m@th\normalfont\normalsize#1}}
\makeatother

\DeclareMathAlphabet{\mathpzc}{OT1}{pzc}{m}{it}

\begin{document}

\title{Identifying Non-Ideal Reaction-Diffusion Systems Unable to Maintain Diffusion Out-of-Equilibrium}

\newcommand\unilu{\affiliation{Complex Systems and Statistical Mechanics, 
Department of Physics and Materials Science, 
University of Luxembourg, 
30 Avenue des Hauts-Fourneaux, L-4362 Esch-sur-Alzette, Luxembourg}}
\newcommand\unipdchem{\affiliation{Department of Chemical Sciences, University of Padova, Via F. Marzolo, 1, I-35131 Padova, Italy}}
\author{Francesco Avanzini}
\email{francesco.avanzini@unipd.it}
\unipdchem
\author{Timur Aslyamov}
\email{timur.aslyamov@uni.lu}
\unilu
\author{Massimiliano Esposito}
\email{massimiliano.esposito@uni.lu}
\unilu


\date{\today}

\begin{abstract}
We develop a general \method{}, based on the construction of a kinetic potential 
serving as a Lyapunov function, 
to establish when diffusion necessarily equilibrates
in non-ideal reaction-diffusion systems
under arbitrary driving by autonomous homogeneous chemostats.
This equilibration of diffusion implies vanishing diffusion currents and homogeneous chemical potentials,
while reactions can remain far from equilibrium.
Using this \method{},
we generalize the results of 
\href{https://pubs.aip.org/aip/jcp/article/161/17/174108/3318625}{J. Chem. Phys. \textbf{161}, 174108 (2024)}
by relaxing some of the underlying assumptions. 
Specifically, we show
that diffusion equilibrates in reaction-diffusion systems
whose chemical reaction network is
either pseudo-detailed balanced, with reaction fluxes controlled by the stoichiometry of reactants and products,
or complex balanced, with reaction fluxes controlled only by the stoichiometry of the reactants.
These different constraints on the reaction fluxes are shown to originate 
from the distinct stoichiometric properties of the two classes of networks.
\end{abstract}

\maketitle



\section{Introduction}
Self-organization of molecules is ubiquitous in nature, giving rise to complex spatio-temporal structures.
This phenomenon has gained renewed attention following the discovery of biomolecular condensates in cells~\cite{Banani:2017condensates, Lyon2021, aierken2026roadmapcondensatescellbiology}:
membrane-less compartments,
resulting from liquid-liquid phase separation and serving a wide range of biological functions~\cite{Klosin2020aa, Castellana2014aa, Peeples2020, Frottin2019aa}.
Indeed, understanding the physical principles underlying the formation and regulation of such condensates remains challenging.
In contrast to equilibrium self-organization~\cite{cahn1958free},
which is governed solely by molecular interactions and diffusion,
biomolecular condensates arise in complex environments,
where their constituents are involved in chemical reactions that are continuously driven out of equilibrium.

Recently, significant efforts have been devoted to elucidating these physical principles using complementary approaches.
Microscopic, particle-based approaches
have been used to investigate molecular-scale effects,
such as
crowding~\cite{Fries2025_crowded},
the frequency of molecular encounters~\cite{Fries2025_encounter},
and interfacial phenomena~\cite{Cho2023_interfacial, Cho2025_interfacial}.
Mesoscopic, mean-field approaches
have been extensively developed to investigate a wide range of phenomena~\cite{weber2019physics, zwicker2022intertwined}.
These include, but are not limited to,
the emergence of specific spatial concentration profiles~\cite{demarchi2023enzyme, glotzer1995reaction, brauns2020phase, haas2021turing},
and their regulation by chemical reactions~\cite{Kirschbaum2021, bauermann2022energy, Rossetto2025_dimerization, Ziethen2023_nucleations}
and molecular interactions~\cite{Menou2023_interactions};
the nature of instabilities of the homogeneous phase~\cite{aslyamov2023nonideal};
the factors controlling coarsening of self-organized structures~\cite{Weyer2023_coarsening, Bauermann2025_ripening, rashofer2026nonequilibriumphasecoexistenceconserved, rashofer2026localcompositioncontrolspattern};
and the role of condensates as microreactors regulating chemical reactions~\cite{Laha2024_regulation}.

These advances have been instrumental in shaping our current understanding of biomolecular condensates.
Nevertheless, many existing approaches rely on phenomenological descriptions
that do not systematically enforce thermodynamic consistency,
thereby limiting the identification of general principles governing nonequilibrium self-organization.
To address this limitation,
we developed in a previous work~\cite{avanzini2024}
a thermodynamically consistent mean-field theory for open, non-ideal reaction-diffusion systems,
by combining thermodynamic frameworks for non-ideal chemical reaction networks~\cite{Othmer1976, Avanzini2021}
and ideal reaction-diffusion systems~\cite{Falasco2018a, Avanzini2019a}.
Within this framework, we identified classes of chemical reaction networks
(i.e., pseudo-detailed balanced and complex balanced)
for which diffusion cannot be sustained out of equilibrium, 
irrespective of the strength of the nonequilibrium driving 
by autonomous homogeneous chemostats.
Eventually, diffusion currents vanish and all chemical potentials become homogeneous 
even though reactions can remain out of equilibrium.
These results were derived
for arbitrary molecular interactions
and under specific assumptions on the reaction fluxes,
although the underlying physical mechanism was expected to be more general.
This broader expectation, however, appeared to clash with particle-based simulations
of reaction-diffusion systems with pseudo-detailed balanced networks,
where diffusion was observed to remain out of equilibrium~\cite{Fries2025_crowded, Cho2025_interfacial}.

To understand this apparent contradiction, 
we develop a general \method{} 
whose central ingredient is the construction of a kinetic potential acting as a Lyapunov function.
The \method{} establishes when diffusion necessarily equilibrates
and is applied to both pseudo-detailed balanced and complex balanced networks.
This allows us to identify the fundamental conditions that reaction fluxes must satisfy for diffusion to equilibrate
within these two classes of networks.
On the one hand,
reaction fluxes in pseudo-detailed balanced chemical reaction networks
must be controlled by the stoichiometry of potentially both reactants and products,
as formally specified by Eq.~\eqref{eq:flux_ass1}.
In Ref.~\cite{avanzini2024}, we instead considered reaction fluxes controlled by the stoichiometry of the reactants only.
On the other hand,
reaction fluxes in complex balanced chemical reaction networks
need only be controlled by the stoichiometry of the reactants,
as formally specified by Eq.~\eqref{eq:flux_ass2}.
In Ref.~\cite{avanzini2024}, we instead considered Arrhenius-like reaction fluxes only.
When reaction fluxes do not satisfy these conditions,
as in Refs.~\cite{Fries2025_crowded, Cho2025_interfacial},
diffusion can instead be maintained out of equilibrium.
Furthermore, the \method{} reveals the conditions leading to
a novel lower bound on the entropy production rate due to reactions.

The paper is organized as follows. 
Section~\ref{sec:setup} summarizes the main setup for the mean-field dynamic and thermodynamic description 
of reaction-diffusion systems.
Section~\ref{sec:classCRNs} introduces the two classes of chemical reaction networks based on their topological properties.
Section~\ref{sec:eq_diff} derives our results.
Section~\ref{sec:examples} illustrates the conditions leading to the equilibration of diffusion through two specific examples.
Our results
are then discussed and compared with other results in the literature in Sec.~\ref{sec:discussion}.



\section{Basic Setup\label{sec:setup}}

We consider open, non-ideal reaction-diffusion (RD) systems in solution
at constant temperature $T$ and volume $V$. 
Chemical species (labeled $\spe\in\setspe$) 
interact, react, diffuse, and can be exchanged with external chemostats.
If exchanged, they are said to be chemostatted (labeled $\spey\in\setspey\subseteq\setspe$).
Otherwise, they are said to be internal (labeled $\spex\in\setspex = \setspe \setminus \setspey$).
Elementary reactions (labeled $\rct\in\setrct$) are reversible:
if $\rct\in\setrct$ denotes a forward reaction, 
then $-\rct\in\setrct$ denotes its backward counterpart.
Each pair $\pm\rct$ of reactions is represented by the equation
\begin{equation}\small
\sum_{\spex}\stoc{}{\spex}\,\chemx  +
\sum_{\spey}\stoc{}{\spey}\,\chemy
\ch{ <=>[$\rct$][$-\rct$]}
\sum_{\spex}\stoc{-}{\spex}\,\chemx  +
\sum_{\spey}\stoc{-}{\spey}\,\chemy
\,,
\label{eq:cr}
\end{equation}
where $\chem$ denotes the chemical symbol of species $\spe$,
while $\stoc{}{\spe}$ (resp. $\stoc{-}{\spe}$) denotes its stoichiometric coefficient in reaction $\rct$ (resp. $-\rct$).
The net variation in the molecule number of species $\spe$ in reaction $\rct$ is given by
\begin{equation}
\matSi{}=\stoc{-}{\spe} - \stoc{}{\spe}\,,
\label{eq:matS}
\end{equation}
which is the $(\spe,\rct)$ entry of the stoichiometric matrix $\matS$. 
Equation~\eqref{eq:cr} defines a hypergraph, known as a chemical reaction network (CRN)~\cite{Klamt2009},
where species are mapped to nodes and reactions to edges.

We focus on open, non-ideal RD systems in the macroscopic limit~\cite{falasco2024macroscopic}.
In this limit, the abundances of the chemical species 
at every point in space $\pos\in V$ are specified by the concentration fields $\conv = (\dots,\conc,\dots)$,
which evolve according to the following (deterministic) RD equation:
\begin{equation}
\pt \conc = 
-\ds\cdot\dcurr
+\sum_{\rct>0}\matSi{}\,\rcurr{}
+ \ecurr\,,
\label{eq:rdeq}
\end{equation}
where 
$\ecurr$ is the rate at which species $\spe$ is exchanged with the corresponding chemostat;
$\rcurr{}$ is the net current of reaction $\rct>0$ reading 
\begin{equation}
\rcurr{} = \rflux{} - \rflux{-}\,,
\end{equation}
where $\rflux{}$ and $\rflux{-}$ are the fluxes of the forward and backward reactions, respectively;
$\dcurr$ is the diffusion current of species $\spe$. 

RD systems are dissipative systems that obey thermodynamic constraints.
When all degrees of freedom other than the concentration fields are equilibrated,
reaction fluxes $\rflux{\pm}$ satisfy the local detailed balance condition~\cite{postmodernthermo2023},
reading 
\begin{equation}\small
RT\ln\frac{\rflux{}}{\rflux{-}} = 
-\Big(
\sum_{\spex}\cpx\matSx{}
+\sum_{\spey}\cpy\matSy{}
\Big)\,,
\label{eq:ldb}
\end{equation}
while, within the linear regime, the diffusion current $\dcurr$ of each species satisfies 
\begin{equation}
\dcurr = -\sum_{\spee}\matOij\ds\cpg{\spee}\,, 
\label{eq:dcurr}
\end{equation}
with $\{\matOij\}$ being the entries of the positive-definite Onsager matrix $\matO$~\cite{Groot1984}.
Here, $\cp = \fd{\free}$ is the chemical potential of species~$\spe$,
while $\free$ is the nonequilibrium Helmholtz free energy of the RD system.
Correspondingly,  
the second law of thermodynamics can be formulated as 
a balance equation for the Helmholtz free energy,
which (according to the RD equation~\eqref{eq:rdeq}) reads
\begin{equation}
\dt\free = - T (\eprrct + \eprdff) + \chwrk
\label{eq:2law}
\,.
\end{equation}
On the one hand, 
\begin{subequations}\label{eq:epr}
\begin{align}
T\eprrct &\equiv -\sum_{\spe} \sum_{\rct>0}\int_V\dd\pos\, \cp \matSi{}\rcurr{} \geq 0\label{eq:eprrct1}\\
T\eprdff &\equiv - \sum_{\spe}\int_V\dd\pos\, \ds\cp\cdot\dcurr \geq 0  \label{eq:eprdff1}
\end{align}%
\end{subequations} 
are the entropy production rates due to reactions and diffusion, respectively, 
quantifying dissipation.
On the other hand,
\begin{equation}
\chwrk \equiv \sum_{\spey} \int_V\dd\pos\, \cpy\, \ecurry
\end{equation}
is the chemical work rate,
quantifying the free energy exchanged with the chemostats 
and driving RD systems out of equilibrium.

From a thermodynamic perspective,
a steady state is characterized by time-independent concentration fields $\convvss$ and thus by
$\dt\freess = 0$.
In a nonequilibrium steady state, reactions and diffusion may nevertheless remain out of equilibrium and dissipate,
with $\eprrctss > 0$ and $\eprdffss > 0$.
According to the second law in Eq.~\eqref{eq:2law},
such dissipation must be sustained by a continuous supply of free energy from the chemostats,
i.e., $\chwrkss = T(\eprrctss + \eprdffss) > 0$.
Equilibrium is instead a particular steady state $\convveq$ in which
both reactions and diffusion have equilibrated and therefore do not dissipate:
$\eprrcteq = 0$ and $\eprdffeq = 0$.
According to the second law in Eq.~\eqref{eq:2law},
the chemical work rate then vanishes, i.e., $\chwrkeq = 0$.
We are interested here in particular nonequilibrium steady states in which
only diffusion has equilibrated, while reactions remain out of equilibrium and dissipate:
$\eprdffss = 0$ and $\eprrctss > 0$.
According to Eqs.~\eqref{eq:eprdff1} and~\eqref{eq:dcurr},
all diffusion currents vanish, $\dcurrss = 0$,
and all chemical potentials become homogeneous, $\ds\cpss = 0$,
while the reaction currents can remain nonzero, $\rcurrss{} \neq 0$.
According to the second law in Eq.~\eqref{eq:2law},
the resulting reaction dissipation is sustained by the chemical work performed by the chemostats,
i.e., $\chwrkss = T\eprrctss > 0$.

\section{Topological Classification of CRNs\label{sec:classCRNs}}
Here we introduce \cmark{two} classes of CRNs,
based on the topological properties of \textit{cycles}, \textit{complexes}, and \textit{\pathways}.

\subsection{Topological Properties}
A cycle is a sequence of reactions that, upon completion, does not change the molecule number of any internal species.
It is mathematically defined as a right-null vector $\cyclev = (\dots,\cycle,\dots)$
of the stoichiometric matrix restricted to the internal species:
\begin{equation}
\sum_{\rct>0}\matSx{}\cycle{}=0 \,.
\label{eq:phi}
\end{equation}
A cycle may, however, change the molecule numbers of the chemostatted species, 
i.e., $\sum_{\rct>0}\matSy{}\cycle{}\neq0$,
thus giving rise to an effective reaction involving only chemostatted species.

Complexes (labeled $\com\in\setcom$) are aggregates of internal species acting as reactants:
$\comp{(\rct)} = \sum_{\spex}\stoc{}{\spex},\chemx$.
Each pair $\pm\rct$ of reactions can thus also be represented by the equation
\begin{equation}\small
\comp{(\rct)}
\ch{ <=>[$\rct$][$-\rct$]}
\comp{(-\rct)}
\,,
\end{equation}
defining a graph, known as the graph of complexes~\cite{Horn1972},
where complexes are mapped to nodes and reactions to edges.
The role of complex $\com$ as a product (positive value) or reactant (negative value) in reaction $\rct$
is specified by
\begin{equation}
\matIe = \dk_{\com,\com(-\rct)} - \dk_{\com, \com(\rct)}
\,,
\label{eq:matI}
\end{equation}
which is the $(\com,\rct)$ entry of the incidence matrix $\matI$
(with $\dk_{\com,\com'}$ being the Kronecker delta).
The graph of complexes provides a coarse-grained representation of the corresponding CRN,
obtained by neglecting the chemostatted species.
As a result,
the incidence matrix and the stoichiometric matrix (restricted to the internal species)
are related by
\begin{equation}
\matSx{} = \sum_\com \matCe{}\, \matIe
\,,
\label{eq:decompositionSX}
\end{equation}
where $\matCe{(\rct)} = \stoc{}{\spex}$ is the $(\spex,\com)$ entry of the composition matrix $\matC$.

\Pathways{} (labeled $\pth\equiv(\com,\com')\in\setpth$) are aggregates of reactions
interconverting the same pair of complexes
($\comp{}$ into $\comp{'}$):
\begin{equation} 
\setrctpth{}\equiv 
\big\{\rct\in \setrct
 \text{ such that } 
 \comp{(\rct)} = \comp{} 
 \text{ and } \comp{(-\rct)} = 
 \comp{'}\big\}
 \,. 
 \end{equation} 
By definition,
if $\rct\in\setrctpth{}$, then $-\rct\in\setrctpth{-}$ with $-\pth\equiv(\com', \com)$.
Furthermore, all reactions $\rct\in\setrctpth{}$ share
the same stoichiometric coefficients and stoichiometric matrix for each internal species:
$\stocx{} = \stocxe{}$, $\stocx{-} = \stocxe{-}$, and $\matSx{} = \matSxe{}$.
However, 
there may exist two reactions $\rct$ and $\rct'$ in $\setrctpth{}$ such that $\matSy{}\neq\matSyy{}$.
In the following, 
we denote by $\pthrct{}$ the \pathway{} associated with reaction $\rct$, 
i.e., $\pth$ such that $\rct\in\setrctpth{}$.

\subsection{Classification of CRNs\label{sub:classCRNs}}
\begin{figure}
    \centering
    \includegraphics[width=0.40\textwidth]{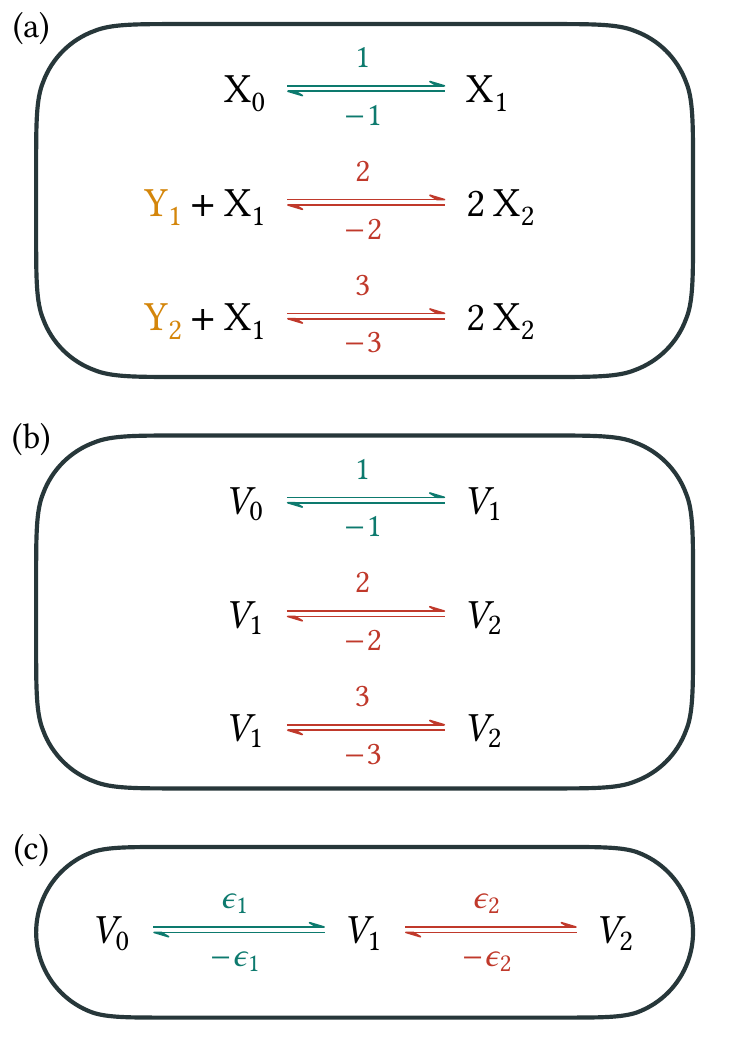}
    \caption{Illustration of a pseudo-detailed balanced CRN.
    (a) A CRN interconverting the 
    internal species $\{\ch{X_0}, \ch{X_1}, \ch{X_2}\}$ 
    and the chemostatted species $\{\ch{Y_1}, \ch{Y_2}\}$ (in amber)
    via the reactions $\{\pm 1, \pm 2, \pm 3\}$.
    (b) Graph of complexes obtained by identifying the complexes 
    $V_0 = \ch{X_0}$, 
    $V_1 = \ch{X_1}$, 
    $V_2 = \ch{2 X_2}$.
    (c) Graph of complexes where the reactions are aggregated in \pathways.
    The reactions $1$ and $-1$ (in teal) correspond to 
    the \pathways{} $\epsilon_1 = (V_0, V_1)$ and $-\epsilon_1 = (V_1, V_0)$,
    respectively. 
    The reactions $\{2,3\}$ and $\{-2,-3\}$ (in red) correspond to 
    the \pathways{} $\epsilon_2 = (V_1, V_2)$ and $-\epsilon_2 = (V_2, V_1)$,
    respectively. 
    Crucially, 
    the cycle corresponding to the sequence of reactions $\{2, -3\}$ (or, equivalently, $\{-2, 3\}$) appearing in (b)
    disappears in (c) after aggregating reactions into the corresponding  \pathways{}.
    \label{fig:ill_pdb}}
\end{figure}
A CRN is said to be \textit{pseudo-detailed balanced} (see Fig.~\ref{fig:ill_pdb}) when
each cycle consists of a sequence of reactions interconverting the same pair of complexes.
Mathematically,
there exists no right-null vector $\cycleev = (\dots,\cyclee,\dots)$ of the stoichiometric matrix 
obtained by grouping reactions belonging to the same \pathway:
\begin{equation}
\sum_{\pth>0}\matSxe{}\, \cyclee = 0\,.
\label{eq:path_balanced}
\end{equation}

\begin{figure}
    \centering
    \includegraphics[width=0.40\textwidth]{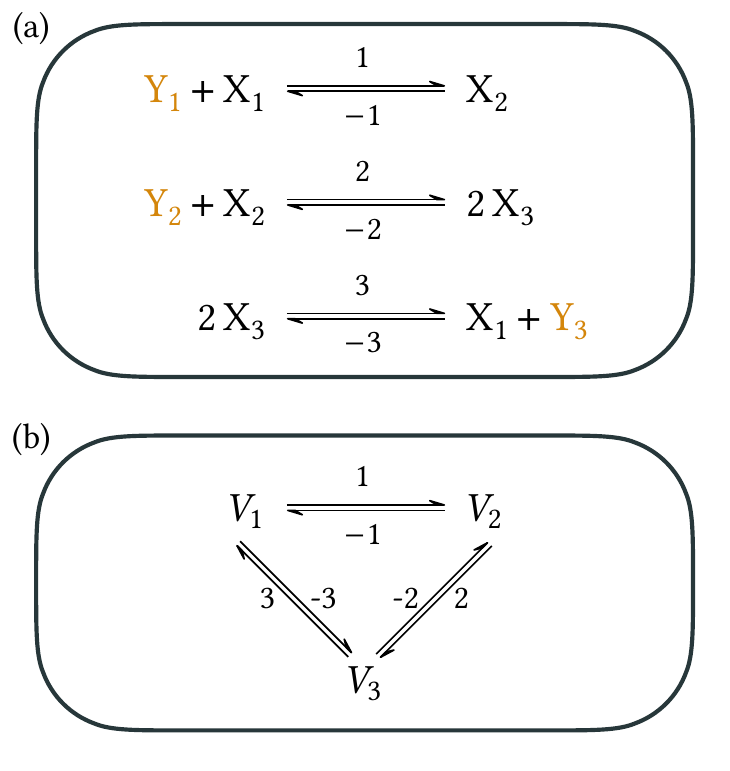}
    \caption{Illustration of a complex balanced CRN.
    (a) A CRN interconverting the 
    internal species $\{\ch{X_1}, \ch{X_2}, \ch{X_3}\}$ 
    and the chemostatted species $\{\ch{Y_1}, \ch{Y_2}, \ch{Y_3}\}$ (in amber)
    via the reactions $\{\pm 1, \pm 2, \pm 3\}$.
    (b) Graph of complexes obtained by identifying the complexes 
    $V_1 = \ch{X_1}$, 
    $V_2 = \ch{X_2}$,
    and $V_3 = \ch{2 X_3}$.
    Crucially, 
    the only cycle of the CRN in (a) corresponding to the sequence of reactions $\{1, 2, 3\}$ (or, equivalently, $\{-1, -2, -3\}$)  
    is the loop featuring the graph of complexes in (b).
    \label{fig:ill_cb}}
\end{figure}
A CRN is said to be (unconditionally) \textit{complex balanced} (see Fig.~\ref{fig:ill_cb}) when 
each cycle corresponds to a loop in the graph of complexes.
Mathematically, each $\cyclev$ is also a right-null vector of the incidence matrix:
\begin{equation}
\sum_{\rct>0}\matIe{}\, \cycle = 0\,.
\label{eq:path_balanced}
\end{equation}
Note that, from Eq.~\eqref{eq:decompositionSX}, 
any right-null vector of the incidence matrix is 
a right-null vector of the stoichiometric matrix restricted to the internal species, i.e., a cycle,
but the converse does not hold.
This difference is encoded in the deficiency
$\deff \equiv \text{dim ker}( \matSX ) - \text{dim ker} (\matI)$,
with $\text{dim ker}(\bullet)$ returning the dimension of the kernel of a matrix.
CRNs are (unconditionally) complex balanced when $\deff = 0$.

\section{Equilibration of Diffusion\label{sec:eq_diff}}
We now introduce a general \method{} to establish when diffusion equilibrates, i.e., $\lim_{t\to\infty} \eprdff = 0$,   
in RD systems with autonomous (i.e., $\pt \cpyy = 0$) homogeneous (i.e., $\ds \cpyy = 0$) chemostatting,
while reactions can remain out of equilibrium, i.e., $\lim_{t\to\infty} \eprrct \neq 0$.
We present the \method{} assuming that the chemostatted species are ideal,
which allows us to treat both $\{\cpyy\}$ and $\{\conyy\}$ as constant parameters,
with $\cpyy = \stcpy + RT \ln \conyy$.
This assumption can nevertheless be relaxed by following the same reasoning as in Ref.~\cite{avanzini2024}.

The \method{} consists of constructing a kinetic potential $\kinfree$ that acts as a Lyapunov function.
Specifically,
the first step consists in identifying concentration-independent coefficients $\{\shx\}$, which 
may encode kinetic properties,
and thereby construct $\kinfree$ as
\begin{equation}
\kinfree = 
\free +
\sum_{\spex} \shx \int_V\dd\pos\,\conx
\,,
\label{eq:kinfree_def}
\end{equation} 
where $\free$ is the nonequilibrium Helmholtz free energy. 
Since $\dt\conyy = 0$ and $\ds \cpyy = 0$,
the balance equation for $\kinfree$ is given by
\begin{equation}
\dt\kinfree = 
- T\kineprrct - T\eprdff 
\,,
\label{eq:kinfree_balance}
\end{equation}
where $\eprdff \geq 0$ is the entropy production rate due to diffusion given in Eq.~\eqref{eq:eprdff1},
while $\kineprrct$ reads
\begin{equation}
T\kineprrct = -\sum_{\spex} \sum_{\rct>0}\int_V\dd\pos\, (\cpx + \shx) \matSx{}\rcurr{}
\,.
\label{eq:kineprrct_gen_def}
\end{equation}
The second step consists in proving that
\begin{equation}
\kineprrct \geq 0
\,.
\label{eq:kineprrct_NN}
\end{equation}
Diffusion equilibrates if these two steps can be achieved. 
Indeed, they imply $\dt\kinfree \leq 0$, which,
together with the fact that $\kinfree$ is lower bounded (since $\free$ must be lower bounded),
implies relaxation to a steady state $\convss$ satisfying
\begin{equation}
\kineprrctss = 0 \text{ }\text{ }\text{ }\text{ and }\text{ }\text{ }\text{ } \eprdffss = 0\,.
\label{eq:crns_eq0}
\end{equation}

This \method{} resembles a standard strategy in thermodynamics:
one modifies a thermodynamic quantity to construct the appropriate thermodynamic potential under the imposed constraints,
whose monotonic decrease identifies the corresponding equilibrium conditions.
Here, the nonequilibrium Helmholtz free energy $\free$ is modified through Eq.~\eqref{eq:kinfree_def}
to construct the potential $\kinfree$.
The added terms effectively incorporate the constraints imposed by the chemostats.
If the coefficients $\{\shx\}$ can be chosen such that Eq.~\eqref{eq:kineprrct_NN} holds,
both terms on the right-hand side of Eq.~\eqref{eq:kinfree_balance} contribute to the monotonic decrease of $\kinfree$.
In particular, persistent diffusion currents, implying $\eprdff > 0$, would continuously decrease $\kinfree$,
which is impossible asymptotically because $\kinfree$ is lower bounded.
Diffusion must therefore equilibrate.
Persistent reaction currents, however, do not necessarily imply $\kineprrct > 0$.
Consequently, $\kineprrct$ can vanish asymptotically while reactions remain out of equilibrium and continue to dissipate.
The analogy with an ordinary thermodynamic potential is nevertheless only partial.
First, $\kinfree$ is not a thermodynamic potential, but rather a kinetic potential:
it is constructed from the nonequilibrium Helmholtz free energy using the coefficients $\{\shx\}$,
which may encode kinetic properties.
Second, of the two non-negative quantities on the right-hand side of Eq.~\eqref{eq:kinfree_balance},
only $\eprdff$ is a physical entropy production rate.
Although $\kineprrct$ acts mathematically as an entropy production rate,
it does not generally coincide with the entropy production rate due to reactions.

We next use this \method{} for two classes of CRNs, namely,
pseudo-detailed balanced CRNs (in Subs.~\ref{sub:pdb})
and complex balanced CRNs (in Subs.~\ref{sub:cb}).
For each class, we identify suitable coefficients $\{\shx\}$ and show that
the corresponding $\kineprrct$ is non-negative.
We use mild class-specific assumptions on the reaction fluxes,
which are introduced and discussed at the beginning of each subsection.
We then conclude (in Subs.~\ref{sec:bound}) by showing how $\kineprrct \geq 0$ can lead to
a novel lower bound on the entropy production rate due to reactions.

\subsection{Pseudo-Detailed Balanced CRNs\label{sub:pdb}}
\assumptionrf 
We assume that the reaction fluxes take the form
\begin{equation}
\rflux{} = 
\rfluxxpth{} \, \rfluxxy{}\,.
\label{eq:flux_ass1}
\end{equation}
Here, 
the contribution of the chemostatted species 
is given by the multiplicative factor $\rfluxxy{}$,
which \--- due to autonomous homogeneous chemostatting \--- is constant
and may be reaction-specific.
This is consistent with
the assumption that chemostatted species are ideal 
and 
with the standard approach in stochastic thermodynamics,
where chemostatted species are treated as sources of nonconservative forces rather than as dynamical species  
(see, e.g., Refs.~\cite{Kirschbaum2021, Cotton2022, zwicker2022intertwined, meyberg2026}).
The contribution of the internal species and their dynamical concentrations (included in $\conv$) 
is given by the factor $\rfluxxpth{}$,
which is not reaction-specific, but depends only on the \pathway{} $\pthrct{}$.

As special case,
Eq.~\eqref{eq:flux_ass1} can describe 
fluxes whose functional dependence on the dynamical concentrations is solely determined by 
the stoichiometric coefficients of the internal species acting as reactants in reaction~$\rct$,
i.e., $\rfluxxpth{} = \rfluxx$,
as already considered in Ref.~\cite{avanzini2024}.
These include, but are not limited to,
Arrhenius-like fluxes, 
which can be written as 
\begin{equation}
\rfluxar{}= 
\AAAp\,
e^{(\sum_{\spex}\cpx\stocx{} 
+ \sum_{\spey}\cpyy\stocy{}) / RT}
\label{eq:arrhenius}
\end{equation}
with $\AAAp$ a constant parameter.

More generally,
Eq.~\eqref{eq:flux_ass1} can also describe
fluxes whose functional dependence on the dynamical concentrations is determined by 
the stoichiometric coefficients of the internal species acting as both reactants and products in reaction $\rct$ 
i.e., $\rfluxxpth{} = \rfluxxx$, 
which were not considered in Ref.~\cite{avanzini2024}.
These include the reaction fluxes derived within transition state theory in Ref.~\cite{Zwicker2025},
which, using our notation, read 
\begin{align}\small
\rfluxts{} = \AAAp 
&\bigg[
\prod_\spex (\conx)^{\stocx{} + \tsq\matSx{} }
\prod_\spey (\conyy)^{\stocy{} + \tsq\matSy{} }
\bigg]
\notag
\\
\times
&\bigg[
e^{-\tsq \frac{\sum_{\spex}\cpx\matSx{} + \sum_{\spey}\cpyy\matSy{}}{RT}}
\bigg]
\,,
\label{eq:flux_zwicker1}
\end{align}
and
\begin{align}\small
\rfluxts{-} = \AAAp 
&\bigg[
\prod_\spex (\conx)^{\stocx{} + \tsq\matSx{} }
\prod_\spey (\conyy)^{\stocy{} + \tsq\matSy{} }
\bigg]
\notag
\\
\times
&\bigg[
e^{-(\tsq - 1) \frac{\sum_{\spex}\cpx\matSx{} + \sum_{\spey}\cpyy\matSy{}}{RT}}
\bigg]
\,,
\label{eq:flux_zwicker2}
\end{align}
provided that the symmetry factor $\tsq \in [0,1]$,
which quantifies whether the transition state is more similar to the reactants ($\tsq < 1/2$) or to the products ($\tsq > 1/2$), 
is the same for all reactions belonging to the same \pathway,
i.e., $\tsq = \tsqq$ for all $\rct\in \setrctpth{}$.
They also include the reaction fluxes used in Ref.~\cite{meyberg2026} for all reactions of the single \pathway{} 
$\ch{X_1 <=> X_2}$,  
which, using our notation, read
\begin{equation}\small
\rfluxts{\pm} = \kconst{\pm} \sqrt{c_{\ch{X_1}}(\pos)c_{\ch{X_2}}(\pos)} e^{\mp (\mu_{\ch{X_2}}(\conv)- \mu_{\ch{X_1}}(\conv)) / 2RT} \,,
\label{eq:flux_speck}
\end{equation}
provided that the transition rates $\{\kconst{\pm}\}$,
satisfying in general $\kconst{}/ \kconst{-} = \exp(-(\sum_{\spey}\cpyy\matSy{}) / RT)$,
are constant, i.e., independent of the dynamical concentrations.
Note that
the fluxes in Eq.~\eqref{eq:flux_speck} with constant transition rates
can be interpreted as a special case of Eqs.~\eqref{eq:flux_zwicker1} and~\eqref{eq:flux_zwicker2} with $\tsq = 1/2$,
but they were introduced independently of Ref.~\cite{Zwicker2025}.

\implementationM
We start by noting that Eq.~\eqref{eq:kineprrct_gen_def} specializes to
\begin{equation}\small
T\eprrcte = -\sum_{\spex} \sum_{\pth>0}\int_V\dd\pos\, (\cpx + \shx) \matSxe{}\pcurr{}
\,,
\label{eq:kineprrct_pdb}
\end{equation}
for pseudo-detailed balanced CRNs, 
using $\sum_{\rct>0} = \sum_{\pth>0}\sum_{\rct\in\setrctpth{}}$,
together with $\matSx{} = \matSxe{}$ for all $\rct\in\setrctpth{}$,
and defining the net current of the  \pathway{} $\pth$ as
\begin{equation}\small
\pcurr{} \equiv \sum_{\rct\in\setrctpth{}} \rcurr{} = \sum_{\rct\in\setrctpth{}}\big\{\rflux{} - \rflux{-}\big\} 
\,.
\label{eq:pcurr_def}
\end{equation}
Hence, a sufficient condition ensuring $\eprrcte\geq 0$ is
that  the current $\pcurr{}$ can be written as the difference between two fluxes~$\pflux{\pm} \geq 0$, namely, 
\begin{equation}\small
\pcurr{} = \pflux{} - \pflux{-}\,,
\label{eq:pcurr3}
\end{equation}
that satisfy the following pseudo local detailed balance condition:
\begin{equation}\small
RT\ln\frac{\pflux{}}{\pflux{-}} = 
- \sum_{\spex}(\cpx + \shx)\matSxe{}\,.
\label{eq:ldb_pb}
\end{equation}
Indeed, under this condition, 
$\eprrcte$ can be rewritten in the manifestly non-negative form
\begin{equation}\small
\eprrcte = R\sum_{\pth>0}\int_V\dd\pos\, 
(\pflux{} - \pflux{-})
\ln\frac{\pflux{}}{\pflux{-}}\geq0
\,.
\label{eq:eprrcte2}
\end{equation}

We now show how to construct such fluxes $\pflux{\pm}$.
To this end, we  write each reaction flux $\rflux{}$ featuring Eq.~\eqref{eq:pcurr_def} as the product of 
its symmetric part $\srr \equiv \sqrt{\rflux{}\rflux{-}}$
and its antisymmetric part $\sqrt{\rflux{}/\rflux{-}}$,
the latter satisfying the local detailed balance condition~\eqref{eq:ldb}.
Hence, 
$\pcurr{}$ can be rewritten as
\begin{equation}\small
\pcurr{} = \aap{}\Big\{ e^{-\frac{\sum_{\spex}\cpx\matSxe{}}{2RT}} - \bbp{} e^{-\frac{\sum_{\spex}\cpx\matSxe{-}}{2RT}}\Big\}\,,
\label{eq:pcurr2}
\end{equation}
where we used $\matSx{} = \matSxe{}$,
while factorizing
\begin{equation}
\aap{} \equiv \sum_{\rct\in\setrctpth{}} \srr e^{-\frac{\sum_{\spey}\cpyy\matSy{}}{2RT}}\geq0\,
\end{equation}
and defining 
\begin{equation}
\bbp{} \equiv \frac{\aap{-}}{\aap{}} = \frac{\sum_{\rct\in\setrctpth{}} \srr{} e^{-\frac{\sum_{\spey}\cpyy\matSy{-}}{2RT}}} {\sum_{\rct\in\setrctpth{}} \srr{} e^{-\frac{\sum_{\spey}\cpyy\matSy{}}{2RT}}}\geq 0.
\label{eq:bbp}
\end{equation}
Note that $\aap{}$ generally depends on the dynamical concentrations,
whereas $\bbp{}$ is constant.
This follows from two conditions.
First, the chemical potentials $\{\cpyy\}$ and the corresponding concentrations $\{\conyy\}$ are homogeneous and constant.
Second, because of the form of the reaction fluxes in Eq.~\eqref{eq:flux_ass1}, 
the dependence of $\srr$ on the dynamical concentrations is uniquely determined by the \pathway,
and not by the specific reaction~$\rct\in\setrctpth{}$:
$\srr \propto \sqrt{ \rfluxxpth{}\rfluxxpth{-}}$.
These contributions can therefore be factorized from the sums in Eq.~\eqref{eq:bbp} 
and  cancel out.

We now use that the CRN is pseudo-detailed balanced.
Since $\matSxe{}$ does not admit right-null vectors, 
we can always introduce constant coefficients $\{\shx\}$ satisfying
\begin{equation}
\ln\bbp{} = \sum_\spex \shx\, \matSxe{} /RT\,.
\label{eq:delta_def}
\end{equation}
Indeed, the set of internal species can be split into two disjoint sets $\setspexi$ and $\setspexd$ such that 
the matrix with entries $\{\matSxe{}\}$ for $\spex\in\setspexi$ and $\setpth\ni\pth>0$ is square and invertible.
Denoting the elements of its inverse by $\{\matSxeI{}\}$, 
the coefficients $\{\shx\}$ can be expressed as
\begin{subequations}
\begin{align}
\shx &= RT \sum_{\pth>0} \ln\bbp{} \, \matSxeI{} \text{ }\text{ }\text{ }\text{ for }\text{ }\text{ }\text{ } \spex\in\setspexi\,,\\
\shx &= 0 \text{ }\text{ }\text{ }\text{ for }\text{ }\text{ }\text{ } \spex\in\setspexd\,.
\end{align}
\label{eq:shift:pdb}%
\end{subequations}
This allows us to rewrite the net current of \pathway{} $\pth$ given in Eq.~\eqref{eq:pcurr2} as in Eq.~\eqref{eq:pcurr3}
by defining
\begin{equation}
\pflux{\pm} \equiv \aap{\abs} e^{\frac{\sum_\spex\shx\matSxe{\abs}}{2RT}}  e^{-\frac{\sum_{\spex}(\cpx + \shx)\matSxe{\pm}}{2RT}}\,.
\end{equation}
These fluxes satisfy the pseudo local detailed balance condition in Eq.~\eqref{eq:ldb_pb}.

\remark
We now emphasize two key aspects of the proof. 
First,
the class of pseudo-detailed balance CRNs ensures 
the existence of the coefficients $\{\shx\}$
defined in Eq.~\eqref{eq:shift:pdb}.
Second, 
the form of the reaction fluxes in Eq.~\eqref{eq:flux_ass1} ensures 
that $\{\bbp{}\}$, defined in Eq.~\eqref{eq:bbp}, are constant,
which \--- together with Eq.~\eqref{eq:shift:pdb} \--- implies that the coefficients $\{\shx\}$ are constant. 
The combined effect of these two aspects is that the net fluxes~$\pflux{\pm}$ of RD systems with pseudo-detailed balanced CRNs
satisfy a pseudo local detailed balance condition 
and, therefore, that the kinetic potential $\kinfree$ in Eq.~\eqref{eq:kinfree_def}
acts as a Lyapunov function.

\subsection{Complex Balanced CRNs\label{sub:cb}}
\assumptionrf 
We assume that the reaction fluxes take the form
\begin{equation}
\rflux{} = 
\rfluxxcom{} \, 
\rfluxxy{}\,.
\label{eq:flux_ass2}
\end{equation}
As for pseudo-detailed balanced CRNs, 
the contribution of the chemostatted species 
is given by the multiplicative factor $\rfluxxy{}$,
which \--- due to autonomous homogeneous chemostatting \--- is constant
and may be reaction-specific as in Subs.~\ref{sub:pdb}.
The contribution of the internal species and their dynamical concentrations (included in $\conv$) 
is given by the factor $\rfluxxcom{}$,
which is not reaction-specific, but depends only on the complex $\comrct{}$.

As special case,
Eq.~\eqref{eq:flux_ass2}, like Eq.~\eqref{eq:flux_ass1}, 
can describe 
fluxes whose functional dependence on the dynamical concentrations is solely determined by 
the stoichiometric coefficients of the internal species acting as reactants in reaction~$\rct$,
i.e., $\rfluxxcom{} = \rfluxx$.
These include, but are not limited to, 
Arrhenius-like fluxes~\eqref{eq:arrhenius} already considered in Ref.~\cite{avanzini2024}.
Unlike Eq.~\eqref{eq:flux_ass1}, Eq.~\eqref{eq:flux_ass2} cannot describe
fluxes whose functional dependence on the dynamical concentrations is determined by 
the stoichiometric coefficients of the internal species acting as both reactants and products in reaction $\rct$,
e.g., the fluxes in Eqs.~\eqref{eq:flux_zwicker1} and~\eqref{eq:flux_zwicker2}. \\

\implementationM
We start by assuming that the kinetic potential $\freecb$ has the same form 
of the known Lyapunov function for ideal complex balanced CRNs~\cite{Anderson2015}, namely, 
$\freecb \equiv \free - \sum_{\spex}\cpxssh\int_V\dd\pos\,\conx$,
where $\convssh$ is a homogeneous steady state.
This corresponds to coefficients $\{\shx\}$ which read 
\begin{equation}
\shx = - \cpxssh \,.
\label{eq:dx_cb}
\end{equation}
Note that these coefficients are defined using the concentrations of an homogeneous steady state, 
but they are independent of the dynamical concentrations $\conv$.
Under this assumption,
Eq.~\eqref{eq:kineprrct_gen_def} specializes to
\begin{equation}\small
T\eprrctc  = -
\sum_{\spex ,\, \rct}
\int_V\dd\pos\big\{
(\cpg{\spex} - \cpgssh{\spex})
\matSx{} \rflux{} \big\}
\,,
\end{equation}
where we used $\sum_{\rct>0}\matSx{}\rcurr{} = \sum_{\rct}\matSx{}\rflux{}$.

We now show that $-\eprrctc\leq 0$. 
To this end, we 
i) use the definition of the stoichiometric matrix in Eq.~\eqref{eq:matS},
ii) multiply and divide by $\exp{(\sum_{\spex}(\cpg{\spex} - \cpgssh{\spex})\stocx{}/{RT})}$,
and iii) use the log inequality $e^\rna(\rnb - \rna) \leq e^\rnb - e^\rna$, 
where equality holds if and only if $\rna = \rnb$ 
(with 
$\rna = \sum_{\spex} (\cpg{\spex} - \cpgssh{\spex}) \stocx{} / RT$ 
and 
$\rnb = \sum_{\spex} (\cpg{\spex} - \cpgssh{\spex}) \stocx{-} / RT$).
We thus obtain
\begin{equation}\small
\begin{split}
&-\frac{\eprrctc}{R}\leq
\int_V\dd\pos \sum_{\rct}
\Big\{
e^{-\frac{\sum_{\spex}(\cpg{\spex} - \cpgssh{\spex})\stocx{}}{RT}}
\rflux{}\\
\times &\Big[
e^{\frac{\sum_{\spex}(\cpx - \cpxssh)\stocx{-}}{RT}} 
-
e^{\frac{\sum_{\spex}(\cpx - \cpxssh)\stocx{}}{RT}}
\Big]\Big\}
\,,
\end{split}
\label{eq:cb_inter}
\end{equation}
where equality holds if and only if $\sum_{\spex} (\cpg{\spex} - \cpgssh{\spex}) \matSx{}=0$.
Note that the integrand in Eq.~\eqref{eq:cb_inter} simplifies to
\begin{equation}
\sum_{\rct}
\Big\{
\rflux{}
\Big[
e^{\frac{\sum_{\spex}(\cpx - \cpxssh)\matSx{}}{RT}} 
-
1
\Big]\Big\}
\,.
\label{eq:integrand_1}
\end{equation}
Applying the change of variables $\rct \to -\rct$ 
to the first term of the sum, together with $\matSx{-} = - \matSx{}$, 
we can rewrite the integrand in Eq.~\eqref{eq:integrand_1} as
\begin{equation}
\sum_{\rct}
\Big[
\rflux{-}e^{-\frac{\sum_{\spex}(\cpx - \cpxssh)\matSx{}}{RT}} 
-
\rflux{}
\Big]
\,.
\label{eq:integrand_2}
\end{equation}
Finally, since $\sum_{\rct} = \sum_{\com}\sum_{\rct}\dk_{\com,\com(\rct)}$, 
$\rflux{} / \rflux{-}$ satisfies the local detailed balance condition~\eqref{eq:ldb}, 
and $\rflux{}$ satisfies Eq.~\eqref{eq:flux_ass2},
the integrand in Eq.~\eqref{eq:integrand_2} becomes 
\begin{equation}\small
\sum_{\com}
\rfluxxcomm
\sum_{\rct}\dk_{\com,\com(\rct)}
\rfluxxy{}
\Big[
e^{\frac{
\sum_{\spex}\cpxssh\matSx{}
+\sum_{\spey}\cpyy\matSy{}
}{RT}}
-
1
\Big] 
\,
\label{eq:integrand_3}
\end{equation}
and vanishes.
Indeed,
we recognize that a homogeneous steady state of a (unconditionally) complex balanced CRN, 
i.e., $\convssh$ such that $\sum_{\rct>0}\matSx{}\,\rcurrssh{}=0$, 
must satisfy 
$\sum_{\rct>0} \matIe\, \rcurrssh{}=0$,
since any right-null vector of the stoichiometric matrix restricted to the internal species 
is also a right-null vector of the incidence matrix. 
Equivalently, a homogeneous steady state must satisfy 
\begin{equation}
\sum_{\rct} \dk_{\com,\com(\rct)}  
\rfluxxy{}
\bigg(1 - 
e^{\frac{
\sum_{\spex}\cpxssh\matSx{}
+\sum_{\spey}\cpyy\matSy{}
}{RT}}
\bigg)  
= 0
\,,
\label{eq:sscb}
\end{equation}
where we used Eqs.~\eqref{eq:matI} and~\eqref{eq:flux_ass2} 
together with the local detailed balance condition~\eqref{eq:ldb}.
Hence, by using Eq.~\eqref{eq:sscb} in Eq.~\eqref{eq:integrand_3},
we obtain that $-\eprrctc \leq 0$ or, equivalently, $\eprrctc \geq 0$.

\remark
We now emphasize one key aspect of the proof. 
The choice of the reaction fluxes in Eq.~\eqref{eq:flux_ass2} ensures that
$\rfluxxcom{}$ depends only on the complex acting as reactant and not on reaction $\rho$.
Thus it can be factorized out of the sum $\sum_{\rct}\dk_{\com,\com(\rct)}$ in Eq.~\eqref{eq:integrand_3},
leading to the emergence of the steady-state condition for complex balanced CRNs given in Eq.~\eqref{eq:sscb}.

\subsection{Lower Bound for the Entropy Production Rate Due to Reactions\label{sec:bound}}
We now show that, in RD systems with autonomous homogeneous chemostatting,
a lower bound on the entropy production rate due to reactions~\eqref{eq:eprrct1}
emerges whenever
$\kinfree$ in Eq.~\eqref{eq:kinfree_def} can be constructed
and the corresponding $\kineprrct$ in Eq.~\eqref{eq:kineprrct_gen_def} is non-negative.
This lower bound can be  
tighter than $\eprrct\ge0$.
To this end, 
we rewrite $\kineprrct$ as
\begin{equation}
T\kineprrct = 
T\eprrct  
+ 
\sum_{\rct>0}
\Big( \sum_{\spex} - \shx \matSx{} + \sum_{\spey} \cpyy \matSy{} \Big)
\rcurravg{}
\,,
\end{equation}
where
we have defined $\rcurravg{} \equiv \int_V\dd\pos\, \rcurr{}$,
added and subtracted $ \sum_{\spey} \cpyy \matSy{} \rcurravg{}$,
and used homogeneous chemostatting (i.e., $\ds \cpyy = 0$) to identify $\eprrct$. 
This, together with $\kineprrct \geq 0$, yields 
\begin{equation}
T\eprrct  
\geq
\sum_{\rct>0}
\Big( \sum_{\spex} \shx \matSx{} - \sum_{\spey} \cpyy \matSy{} \Big)
\rcurravg{}
\,,
\label{eq:eprrct1_LB}
\end{equation}
thus providing a lower bound for $\eprrct$
that is  
tighter than $\eprrct\ge0$
whenever the right-hand side of Eq.~\eqref{eq:eprrct1_LB} is positive.
Although the right-hand side of Eq.~\eqref{eq:eprrct1_LB}
does not appear to admit a general physical interpretation,
for complex balanced CRNs Eq.~\eqref{eq:eprrct1_LB} becomes
\begin{equation}
T\eprrct  
\geq
 \sum_{\rct>0}
\Big( - \sum_{\spex} \cpxssh \matSx{} - \sum_{\spey} \cpyy \matSy{} \Big)
\rcurravg{}
\end{equation}
by using Eq.~\eqref{eq:dx_cb}.
Here, each term in parentheses is the affinity of reaction $\rct$ 
at the homogeneous steady state $\convssh$,
multiplied by the corresponding integrated current evaluated at the instantaneous state $\conv$.

Note that 
the tightness of this bound and the conditions under which it can be saturated remain open questions.

\section{Illustrations\label{sec:examples}}
We now examine, through two illustrative examples,
the conditions on stoichiometry and reaction fluxes
that lead to equilibration of diffusion.

\subsection{Pseudo-Detailed Balanced CRNs}
Consider the following CRN:
\begin{equation}
\begin{split}
\ch{X_1 + Y_1} &\ch{ <=>[$1$][$-1$]} \ch{X_2 + Y_2} \,, \\
\ch{X_1} &\ch{ <=>[$2$][$-2$]} \ch{X_2} \,,
\end{split}
\label{eq:ex1CRN}
\end{equation}
where the internal species $\{\ch{X_1}, \ch{X_2}\}$ and the chemostatted species $\{\ch{Y_1}, \ch{Y_2}\}$
are interconverted through the reactions $\{\pm1, \pm2\}$.
Its stoichiometric matrix reads
\begin{equation}
\matS=
 \kbordermatrix{
    & \color{g}1 &\color{g}2\cr
    \color{g}\ch{X_1}    	&-1 &-1 \cr
    \color{g}\ch{X_2}   	& 1 & 1 \cr
    \color{g}\ch{Y_1}  	&-1 & 0 \cr
    \color{g}\ch{Y_2}   	& 1 & 0 \cr
  }
\label{eq:ex1matS}
\end{equation}
and admits the cycle $\cyclev = (1, -1)^\intercal$.

The corresponding graph of complexes is given by
\begin{equation}
\begin{split}
V_1 &\ch{ <=>[$1$][$-1$]} V_2 \,, \\
V_1 &\ch{ <=>[$2$][$-2$]} V_2 \,,
\end{split}
\end{equation}
upon identifying $V_1 = \ch{X_1}$ and $V_2 = \ch{X_2}$,
and admits only the \pathways{} $\epsilon_1 = (V_1, V_2)$ and $-\epsilon_1 = (V_2, V_1)$.
Consequently, the stoichiometric matrix restricted to the internal species and the \pathway{} $\epsilon_1$ reads
\begin{equation}
 \kbordermatrix{
    & \color{g}\epsilon_1 \cr
    \color{g}\ch{X_1}    	&-1  \cr
    \color{g}\ch{X_2}   	& 1  \cr
  }\,.
\end{equation}
It does not admit any cycle and,
according to Subs.~\ref{sub:classCRNs},
the CRN~\eqref{eq:ex1CRN} is, therefore, pseudo-detailed balanced.
If the reaction fluxes read
\begin{subequations}\small
\begin{align}
\rfluxex{\pm1} = A_{|1|} &
\bigg[ \sqrt{c_{\ch{X_1}} (\pos)c_{\ch{X_2}} (\pos) c_{\ch{Y_1}} c_{\ch{Y_2}} } \bigg] 
\notag
\\
\times
&\bigg[e^{\mp\frac{\mu_{\ch{X_2}}(\conv)- \mu_{\ch{X_1}}(\conv) + \mu_{\ch{Y_2}} - \mu_{\ch{Y_1}}}{2RT}} \bigg]  \,\\
\rfluxex{\pm2} = A_{|2|} &
\bigg[ \sqrt{c_{\ch{X_1}} (\pos)c_{\ch{X_2}} (\pos) } \bigg] 
\notag
\\
\times
&\bigg[e^{\mp\frac{\mu_{\ch{X_2}}(\conv)- \mu_{\ch{X_1}}(\conv)}{2RT}} \bigg] \,
\end{align}
\label{eq:ex1fluxes}%
\end{subequations}
where $A_{|1|}$ and $A_{|2|}$ are constant parameters,
their functional dependence on the dynamical concentrations
is the same for reactions $1$ and $2$ (resp. $-1$ and $-2$)
belonging to \pathway{} $\epsilon_1$ (resp. $-\epsilon_1$).
This implies that the reaction fluxes satisfy Eq.~\eqref{eq:flux_ass1},
and the CRN~\eqref{eq:ex1CRN} eventually relaxes to a steady state in which diffusion has equilibrated,
while reactions can remain out of equilibrium.
Note that the reaction fluxes in Eq.~\eqref{eq:ex1fluxes} are a specialization of the general expression in Eqs.~\eqref{eq:flux_zwicker1} and~\eqref{eq:flux_zwicker2}
to the CRN in Eq.~\eqref{eq:ex1CRN} assuming $\tsq = 1/2$.

We conclude by stressing that diffusion can instead remain out of equilibrium if
$A_{|1|}$ and $A_{|2|}$ depend on the dynamical concentrations in different ways,
such that $A_{|1|}(\conv) \neq A_{|2|}(\conv)$.
Consider, for instance, the case in which $A_{|1|}\propto (c_{\ch{X_1}}(\pos) + c_{\ch{X_2}}(\pos))$,
while $A_{|2|}(\conv)$ is constant.
In this case,
the fluxes of the reactions involving the chemostatted species increase with the total concentration of the internal species,
while the fluxes of the other reactions are independent of it.
Hence, the functional dependence of the reaction fluxes on the dynamical concentrations
is no longer the same for reactions $1$ and $2$ (resp. $-1$ and $-2$)
belonging to \pathway{} $\epsilon_1$ (resp. $-\epsilon_1$),
and Eq.~\eqref{eq:flux_ass1} is not satisfied.

\subsection{Complex Balanced CRNs}
Consider the following CRN:
\begin{equation}
\begin{split}
\ch{Y_1} &\ch{ <=>[$1$][$-1$]} \ch{N} \,, \\
\ch{N} + m \ch{E} &\ch{ <=>[$2$][$-2$]} (m + n) \ch{E} + \ch{W} \,,\\
n\ch{E} + \ch{W} &\ch{ <=>[$3$][$-3$]} \ch{Y_2}\,,
\end{split}
\label{eq:ex2CRN}
\end{equation}
where the internal species $\{\ch{N}, \ch{W}, \ch{E}\}$ and the chemostatted species $\{\ch{Y_1}, \ch{Y_2}\}$
are interconverted through the reactions $\{\pm1, \pm2, \pm3\}$,
with $n$ and $m$ being non-negative integers defining the reaction stoichiometry.
Its stoichiometric matrix reads
\begin{equation}
\matS=
 \kbordermatrix{
    & \color{g}1 &\color{g}2&\color{g}3\cr
    \color{g}\ch{N}    	& 1 &-1 & 0\cr
    \color{g}\ch{W}   	& 0 & 1 & -1 \cr
    \color{g}\ch{E}   	& 0 & n &-n \cr
    \color{g}\ch{Y_1}  	&-1 & 0 & 0 \cr
    \color{g}\ch{Y_2}   & 0 & 0 & 1 \cr
  }
\label{eq:ex2matS}
\end{equation}
and admits the cycle $\cyclev = (1, 1, 1)^\intercal$.

When $n > 0$ and $m = 0$,
the corresponding graph of complexes is simply given by
\begin{equation}
\begin{split}
\emptyset &\ch{ <=>[$1$][$-1$]} V_1 \,, \\
V_1 &\ch{ <=>[$2$][$-2$]} V_2 \,, \\
V_2 &\ch{ <=>[$3$][$-3$]} \emptyset \,,
\end{split}
\end{equation}
upon identifying $V_1 = \ch{N}$ and $V_2 = n \ch{E} + \ch{W}$,
while using $\emptyset$ to represent any complex containing no internal species.
The corresponding incidence matrix
\begin{equation}
\matI=
 \kbordermatrix{
    & \color{g}1 &\color{g}2&\color{g}3\cr
    \color{g}V_1    	& 1 &-1 & 0\cr
    \color{g}V_2   	& 0 & 1 & -1 \cr
    \color{g}\emptyset   &-1 & 0 & 1 \cr
  }
\end{equation}
admits the right-null vector $\cyclev = (1, 1, 1)^\intercal$.
Hence,
according to Subs.~\ref{sub:classCRNs},
the CRN~\eqref{eq:ex2CRN} is complex balanced.
If the reaction fluxes read
\begin{subequations}
\begin{align}
\rfluxex{1} &= A_{|1|} e^{\frac{\mu_{\ch{Y_1}}}{RT}}  \,,\\
\rfluxex{-1} &= A_{|1|} e^{\frac{\mu_{\ch{N}}(\conv)}{RT}}  \,\\
\rfluxex{2} &= A_{|2|} e^{\frac{\mu_{\ch{N}}(\conv)}{RT}}  \,,\\
\rfluxex{-2} &= A_{|2|} e^{\frac{ n \mu_{\ch{E}}(\conv) + \mu_{\ch{W}}(\conv)}{RT}}  \,,\\
\rfluxex{3} &= A_{|3|} e^{\frac{n \mu_{\ch{E}}(\conv) + \mu_{\ch{W}}(\conv)}{RT}} \,,\\
\rfluxex{-3} &= A_{|3|} e^{\frac{\mu_{\ch{Y_2}}}{RT}} \,,
\end{align}
\label{eq:ex2fluxes}%
\end{subequations}
where $A_{|1|}$, $A_{|2|}$, and $A_{|3|}$ are constant parameters,
their functional dependence on the dynamical concentrations
is determined solely by the stoichiometry of the corresponding complex.
It is indeed the same for $\rfluxex{-1}$ and $\rfluxex{2}$, on the one hand, 
and for $\rfluxex{-2}$ and $\rfluxex{3}$, on the other hand.
This implies that the reaction fluxes satisfy Eq.~\eqref{eq:flux_ass2},
and the CRN~\eqref{eq:ex2CRN} eventually relaxes to a steady state in which diffusion has equilibrated,
while reactions can remain out of equilibrium.
Note that the reaction fluxes in Eq.~\eqref{eq:ex2fluxes} are a specialization of the Arrhenius fluxes in Eq.~\eqref{eq:arrhenius}
to the CRN in Eq.~\eqref{eq:ex2CRN}.

When $n > 0$ and $m > 0$,
the corresponding graph of complexes is given by
\begin{equation}
\begin{split}
\emptyset &\ch{ <=>[$1$][$-1$]} V_1 \,, \\
V_3 &\ch{ <=>[$2$][$-2$]} V_4 \,, \\
V_2 &\ch{ <=>[$3$][$-3$]} \emptyset
\end{split}
\end{equation}
upon identifying $V_1 = \ch{N}$, $V_2 = n \ch{E} + \ch{W}$,
$V_3 = \ch{N} + m \ch{E}$, and $V_4 = (m + n) \ch{E} + \ch{W}$,
while using $\emptyset$ to represent any complex containing no internal species.
The corresponding incidence matrix
\begin{equation}
\matI=
 \kbordermatrix{
    & \color{g}1 &\color{g}2&\color{g}3\cr
    \color{g}V_1    	& 1 & 0 & 0\cr
    \color{g}V_2   	& 0 &0 &-1 \cr
    \color{g}V_3   	& 0 & -1 &0\cr
    \color{g}V_4  	& 0 & 1 & 0 \cr
    \color{g}\emptyset   &-1 & 0 & 1 \cr
  }
\end{equation}
does not admit any nontrivial right-null vector and,
according to Subs.~\ref{sub:classCRNs},
the CRN~\eqref{eq:ex2CRN} is not complex balanced.
Hence, the general result derived for complex balanced CRNs no longer applies,
and equilibration of diffusion is not guaranteed.



\section{Discussion and Conclusions\label{sec:discussion}}
We have developed a \method{}, based on the construction of a kinetic potential acting as a Lyapunov function, 
to establish when diffusion equilibrates in RD systems with autonomous homogeneous chemostatting.
We have applied this \method{} to show that diffusion equilibrates in systems 
with either pseudo-detailed balanced or complex balanced CRNs. 
When diffusion equilibrates, according to Eq.~\eqref{eq:eprdff1}, 
the chemical potentials become homogeneous, $\ds\cpss=0$, 
and diffusion currents vanish, $\dcurrss=0$. 

This has been proven under the assumption that reaction fluxes satisfy mild class-specific constraints.
In particular, they must satisfy Eqs.~\eqref{eq:flux_ass1} and~\eqref{eq:flux_ass2}
for pseudo-detailed balanced and complex balanced CRNs, respectively.
Notably, the constraints imposed on the reaction fluxes for complex balanced CRNs 
are more restrictive than those for pseudo-detailed balanced CRNs:
all fluxes satisfying Eq.~\eqref{eq:flux_ass2} also satisfy Eq.~\eqref{eq:flux_ass1}, 
whereas the converse does not generally hold.
This difference originates from the different stoichiometric properties of the two classes
underpinning the proofs presented in Sec.~\ref{sec:eq_diff}.

In pseudo-detailed balanced CRNs,
the crucial ingredient of the proof
is that $\{\bbp{}\}$ are constant.
According to Eq.~\eqref{eq:bbp},
this requires that the terms depending on the dynamical concentrations,
entering the sums in both the numerator and denominator of each $\bbp{}$
through $\{\srr{}\}$ with ${\rct\in\setrctpth{}}$,
cancel out.
Therefore, this condition only allows for reaction fluxes whose dependence on the dynamical concentrations
is determined solely by the \pathway\ $\pth$ and not by the specific reaction $\rct$.

In complex balanced CRNs,
the crucial ingredient of the proof
is that the steady-state condition $\sum_{\rct>0} \matIe\, \rcurrssh{}=0$
can be rewritten as in Eq.~\eqref{eq:sscb},
where the functional dependence on the concentrations appears only through the exponential terms.
This rewriting requires factorizing out
the fluxes of all reactions having the complex $\com$ as a reactant.
Therefore, it only allows for reaction fluxes whose dependence on the dynamical concentrations
is determined solely by the complex $\com$ and not by the specific reaction $\rct$.
This directly yields the steady-state condition in Eq.~\eqref{eq:integrand_3}.

When the reaction fluxes do not satisfy Eq.~\eqref{eq:flux_ass1} for pseudo-detailed balanced CRNs
or Eq.~\eqref{eq:flux_ass2} for complex balanced CRNs,
diffusion can be maintained out of equilibrium
even under homogeneous chemostatting,
and the chemical potentials of the internal species become non-homogeneous at steady state.
This has, for instance, been observed in the illustration presented in Ref.~\cite{meyberg2026},
where a pseudo-detailed balanced CRN with a single \pathway\ was considered within a mean-field approach.
There, different reactions within the same \pathway\ depend on the local total density in different ways.
Specifically, the transition rates $\{\kconst{\pm}\}$ entering the fluxes in Eq.~\eqref{eq:flux_speck}
for reactions involving chemostatted species
increase linearly with the total concentration $\convtot$ of internal species:
$\kconst{\pm} \propto (1 + \zeta \convtot)$,
where the parameter $\zeta$ controls the dependence on the total concentration.
Conversely,
the transition rates $\{\kconst{\pm}\}$ entering the fluxes in Eq.~\eqref{eq:flux_speck}
for reactions not involving chemostatted species remain constant.
As a result,
the concentration dependence of the reaction fluxes is not determined solely by the \pathway.
Consequently,
Eq.~\eqref{eq:flux_ass1} is not satisfied
and the steady-state chemical potentials of the internal species are non-homogeneous.
A similar behavior has also been observed in Refs.~\cite{Berthin2025, Fries2025_crowded},
where a pseudo-detailed balanced CRN with a single \pathway\ was considered within a particle-based approach
under similar assumptions:
reactions involving chemostatted species dominate in the dense phase,
while reactions not involving chemostatted species dominate in the dilute phase.

Beyond pseudo-detailed balanced and complex balanced CRNs, 
a general procedure to construct $\kinfree$ whose corresponding $\kineprrct$ is non-negative is still lacking. 
Whether such a general procedure exists, or whether the construction is inherently network-specific, remains an open question that we leave to future work.

\section*{Acknowledgments}
FA is supported by the project P-
DiSC\#BIRD2023-UNIPD funded by the Department of Chemical Sciences of the University of Padova (Italy).
T.A. is supported by project ThermoElectroChem (C23/MS/18060819) funded by the Fonds National de la Recherche (FNR), Luxembourg.
M.E. is supported by the ERC Advanced Grant EOS (No. 101265122).
FA and ME would like to thank CECAM for organizing the workshop ``Multiscale modeling of chemically active mixtures''
held at CECAM-HQ-EPFL in 2025,
which provided valuable discussions and insights for this work.

%
%


\bibliography{biblio}
\end{document}